\documentclass[%
 reprint,
superscriptaddress,
bibnotes,
amsmath,amssymb,
aps,
]{revtex4-2}

\usepackage{layouts}
\usepackage[utf8]{inputenc}
\usepackage[usenames,dvipsnames]{color}

\usepackage{amsmath}
\usepackage{amsfonts}
\usepackage{amssymb}
\usepackage[pdftex]{graphicx}
\usepackage{tabularx}
\usepackage{epstopdf}
\usepackage{xr}
\usepackage[pdftex]{hyperref}
\hypersetup{a4paper,
pdftitle={Motility-induced crystallization and rotating crystallites} 
pdfauthor={Max Holl, Alina Steinberg and Uwe Thiele}, 
pdfproducer={lateX},
pdfview=FitV, 
pdfstartview=FitB,
linkcolor=blue, 
citecolor=blue, 
urlcolor=red, 
breaklinks=true, 
colorlinks=true,
citebordercolor=0 0 0, 
filebordercolor=0 0 0,
linkbordercolor=0 0 0,
menubordercolor=0 0 0,
urlbordercolor=0 0 0,
pdfhighlight=/I,
pdfborder=0 0 0, 
bookmarksopen=true,
bookmarksnumbered=true
}
\usepackage[capitalise]{cleveref}
\usepackage[usenames,dvipsnames]{color}
\usepackage[normalem]{ulem}

\usepackage{prelim2e}\usepackage[none,bottom]{draftcopy}
\draftcopyName{preprint / }{1.2} 

\newcommand{\dt}{\partial_t}
\renewcommand{\vec}{\mathbf}
\newcommand{\circlenumber}[1]{\raisebox{.5pt}{\textcircled{\raisebox{-.9pt} {#1}}}}

\begin{document}
\title{Motility-induced crystallization and rotating crystallites}
\author{Max Philipp Holl}
\email{max.holl@aalto.fi}
\thanks{ORCID ID: 0000-0001-6451-9723}
\affiliation{Department of Chemistry and Materials Science, Aalto University, P.O. Box 16100, FI-00076 Aalto, Finland}
\affiliation{Academy of Finland Center of Excellence in Life-Inspired Hybrid Materials (LIBER), Aalto University, P.O. Box 16100, FI-00076 Aalto, Finland}
\affiliation{Institute of Theoretical Physics, University of M\"unster, Wilhelm-Klemm-Str.\ 9, 48149 M\"unster, Germany}

\author{Alina Barbara Steinberg}
\email{a\_stei52@uni-muenster.de}
\thanks{ORCID ID: 0000-0001-6598-9700; MH and AS contributed equally to this work.}
\affiliation{Institute of Theoretical Physics, University of M\"unster, Wilhelm-Klemm-Str.\ 9, 48149 M\"unster, Germany}

\author{Uwe Thiele}
\email{u.thiele@uni-muenster.de}
\homepage{http://www.uwethiele.de}
\thanks{ORCID ID: 0000-0001-7989-9271}
\affiliation{Institute of Theoretical Physics, University of M\"unster, Wilhelm-Klemm-Str.\ 9, 48149 M\"unster, Germany}
\affiliation{Center for Nonlinear Science (CeNoS), University of M\"unster, Corrensstr.\ 2, 48149 M\"unster, Germany}
\affiliation{Center for Multiscale Theory and Computation (CMTC), University of M\"unster, Corrensstr.\ 40, 48149 M\"unster, Germany}
\begin{abstract}
Active soft matter frequently shows motility-induced phase separation (MIPS), where self-propelled particles condensate into clusters with an inner liquid-like structure. Such activity may also result in motility-induced crystallization (MIC) into clusters with an inner crystalline structure. We present a higher-order active Phase-Field-Crystal (PFC) model and employ it to study the interplay of passive (i.e., thermodynamic) and active (i.e., motility-induced) condensation and crystallization. Morphological phase diagrams indicate the various occurring phase coexistences and transitions, e.g., the destruction of passive clusters by density-independent activity and the creation of such clusters by a density-dependent activity. Finally, rotating crystallites are analyzed in some detail.
\end{abstract}

\maketitle

Motility-induced phase separation (MIPS) is an important focus of attention in the physics of active soft matter: when the driving activity of self-propelled particles crosses a critical density-dependent threshold the particles phase-separate into dense and dilute fluid phases. In other words they accumulate into high-density liquid-like clusters that coexist with a low-density gas-like phase \cite{Rama2010arcmp,RBEL2012epjt,MJRL2013rmp,CaTa2015arcmp,Mare2016epjt,GWSS2020jpcm,SWWG2020nrp}. Experimental observations \cite{MJRL2013rmp,BBKL2013prl,BDLR2016rmp}, particle-based simulations \cite{FMMT2012prl,MSAC2013prl,PaYM2017pre,CrPS2019pre,CaMP2020prl}, and continuum models mainly based on nonvariational amendments \cite{WTSA2014nc,SBML2014prl,SMBL2015jcp,RaBZ2019epje,BiWi2020jpcm,BiWi2020prr} of the Cahn-Hilliard model for phase decomposition \cite{Bray1994ap} all yield consistent results for the existence and properties of such clusters \cite{BiWi2020jpcm}. 

However, beside these MIPS-produced clusters of liquid-like inner structure, experiments and simulations also increasingly report the existence of active (or ``living'') crystallites, i.e., resting, traveling or rotating particle clusters with an inner crystalline structure \cite{TCPY2012prl,BBKL2013prl,MSAC2013prl,PSSP2013s,PeWL2015prl,BDLR2016rmp,ZoSt2016jpcm,PeLi2018njp, GTDY2018nc} that one might also see as active solids \cite{HaLi2014prl,HBCD2024prl,FTDH2013prl,MaRa2019prl} of finite size. Frequently, they are modelled using active Phase-Field-Crystal (active PFC) models \cite{MeLo2013prl, MeOL2014pre,ChGT2016el,AlPV2016njp,OpGT2018pre,PVWL2018pre,OKGT2020c,OKGT2021pre,HAGK2021ijam}, i.e., active equivalents of the passive PFC model \cite{EKHG2002prl,ELWG2012ap,TARG2013pre} -- a versatile microscopic field theory for colloidal crystallization that itself is a local approximation of a Dynamical Density Functional Theory (DDFT) \cite{ARRS2019pre,VrLW2020ap} and has the form of a conserved Swift-Hohenberg equation \cite{TARG2013pre}.

Although the active PFC model well captures the transition from resting to traveling space-filling crystalline states \cite{MeLo2013prl}, and also predicts the existence of active crystallites in the form of localized states \cite{OpGT2018pre,OKGT2021pre,HAGK2021ijam} it suffers from two major shortcomings: First, it only describes the fluid-solid phase behavior for a single fluid phase (that is normally interpreted as liquid), but fails to capture the liquid-gas transition. Therefore the model is not able to capture the interplay of condensation and crystallization. Second, it does not describe proper \emph{motility-induced crystallization} (MIC) where activity induces crystallization. Instead, with the standard active PFC model an increase in activity tends to destroy preexisting crystalline states \cite{MeLo2013prl,OKGT2020c,OKGT2021pre}. In other words, the crystallization of passive (i.e., thermodynamic) crystalline clusters described by a passive PFC model \cite{TARG2013pre,SAKR2018njp} is subdued by incorporating activity that also results in the transition from resting to moving (and oscillating) crystals \cite{OKGT2020c}.

\begin{figure}[t]
\centering
\includegraphics[width=\hsize]{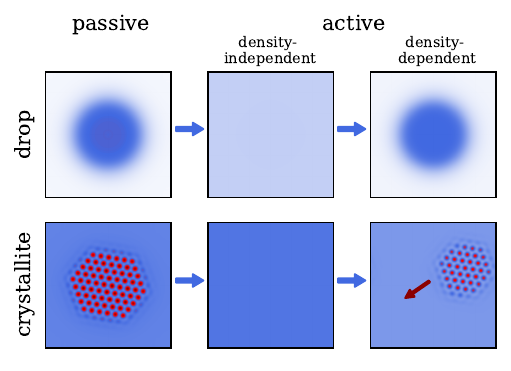}
\caption{A schematic illustration of the destruction of passive (thermodynamic) drops (top) and crystallites (bottom) by an increase of activity from zero (left to center), and the subsequent re-creation of both cluster types by increasing a density-dependent activity.}
\label{fig:cluster}
\end{figure}

Here, we present an extended active PFC model that mitigates both shortcomings: It builds on a passive higher-order PFC model that faithfully captures the basic phase diagram featuring gas, liquid and solid phases including gas-liquid-solid coexistence \cite{WLHD2020prm}. In consequence, it captures the basic dynamics of liquid-gas phase separation as well as of crystallization from either the liquid or the gas phase. As a result of incorporating a density-dependent activity, the developed active model then captures the destruction of passive clusters as the earlier active PFC model \cite{MeLo2013prl} as well as MIC \textit{and} MIPS as schematically illustrated in Fig.~\ref{fig:cluster}. In the following, we present the model, analyze its phase behavior in passive (thermodynamic) and active cases thereby focusing on the destruction of passive clusters (drops and crystallites) and the emergence of active ones. Finally, we analyze the emerging rotating crystallites.

\begin{figure}
\centering
\includegraphics[width=\hsize]{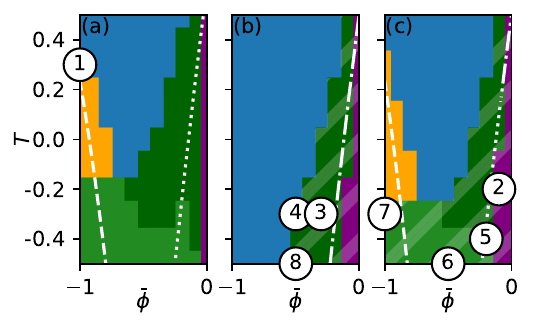}
\caption{Morphological phase diagrams for the extended PFC model \eqref{eq:dtphi} and \eqref{eq:dtP} in the plane spanned by the mean density $\bar\phi$ and the temperature $T$. Shown are (a) the passive case ($v=0$) and two active cases, namely (b) $v_0=1,\, \zeta=0$, and (c) $v_0=1,\,\zeta=-0.5$. Shown is the vicinity of the triple point [at $(\bar\phi, T)\approx (-0.5,0)$] in the passive case. Time simulations are employed for finite domain size $L_x\times L_y = 100\times100$. Examples of occurring states at numbered loci \circlenumber{1}- \circlenumber{8} are given in Fig.~\ref{fig:solutions}. Blue indicates uniform states (gas or liquid), while orange marks liquid-gas coexistence. Dark and light green represent coexistence of a crystalline phase with liquid and gas state, respectively. The latter may also include finite-size realizations of three-phase coexistence. Finally, dark purple indicates domain-filling crystalline states. In (a) all states are at rest while in (b) and (c) traveling states are shown hatched. The white dashed and dotted lines give linear stability thresholds for a Cahn-Hilliard and a conserved-Turing instability related to spinodals for phase separation and crystallization, respectively. The dot-dashed line indicates a conserved-wave instability related to the direct crystallization into traveling crystals. The full set of parameters is given in Tab.~I of the SM. The employed parameter increments for $\bar\phi$ and $T$ are both $0.1$.}
\label{fig:morphological_PDs}
\end{figure}

To model the interplay of active and passive phase transitions we combine two mean field approaches: On the one hand, we use the higher-order passive PFC model for gas-liquid-solid transitions by Wang et al.~\cite{WLHD2020prm}. Its nondimensional kinetic equation corresponds to the gradient dynamics \cite{ELWG2012ap}
\begin{align}
\dt\phi &= \nabla^2\frac{\delta\mathcal{F}[\phi]}{\delta\phi}
\label{eq:passive}
\end{align}
for the conserved order parameter field $\phi(\vec{x},t)$ (a scaled and shifted particle density). In contrast to the standard PFC model \cite{EKHG2002prl,ELWG2012ap,TARG2013pre,ARRS2019pre,VrLW2020ap} the underlying free energy functional $\mathcal{F}[\phi]$ takes into account contributions from three- and four-point direct correlation functions \cite{WLHD2020prm}. The detailed expression is reproduced in section~A of the Supplemental Material (SM) and provides a $\delta\mathcal{F}/\delta\phi$ in \eqref{eq:passive} that contains derivatives up to sixth order. This amended PFC model captures phase transitions between all three phases: the liquid-gas transition between low- and high-density uniform phases as well as liquid-solid and gas-solid transitions between uniform and spatially periodic (crystalline) phases. A typical phase diagram in the vicinity of the triple point (three-phase coexistence) is given in Fig.~\ref{fig:morphological_PDs}~(a). It is obtained by combining direct time simulations of Eq.~\eqref{eq:passive} and linear stability analyses of uniform states (cf.~\cite{HoAT2021jpcm}).

On the other hand, activity resulting from self-propulsion is incorporated following the approach of \cite{MeLo2013prl,MeOL2014pre}. In particular, the passive dynamics of the density \eqref{eq:passive} is coupled to the dynamics of a polarization field $\vec{P}$ that indicates the local strength and direction of polar order related to the drive of the active particles. The resulting dynamic model is
\begin{align}
\dt\phi &= \nabla^2\frac{\delta\mathcal{F}[\phi, \vec{P}]}{\delta\phi} - \nabla\cdot [v(\phi)\vec{P}\label{eq:dtphi}],\\
\dt\vec{P} &= \nabla^2 \frac{\delta\mathcal{F}[\phi, \vec{P}]}{\delta\vec{P}} - D_{r}\frac{\delta\mathcal{F}[\phi, \vec{P}]}{\delta\vec{P}} - \nabla[\alpha v(\phi)\phi].\label{eq:dtP}
\end{align}
The full expression for the density- and polarization-dependent energy $\mathcal{F}[\phi, \vec{P}]$ is given in section~A of the SM. The first and second term on the right hand side of Eq.~\eqref{eq:dtP} correspond to translational and rotational diffusion, respectively. Beside these terms that represent respective conserved and nonconserved gradient dynamics, there are the active coupling terms (the respective final term of Eqs.~\eqref{eq:dtphi} and \eqref{eq:dtP}). Their nonreciprocal character breaks the overall gradient dynamics structure, i.e., represents sustained non-equilibrium influences like chemo-mechanical driving. They are ultimately responsible for all occurring moving and oscillating states.

The strength $v$ of the coupling between the density and polarization represents the self-propulsion speed of the active particles that is often assumed constant \cite{MeLo2013prl,OpGT2018pre}. In contrast, here, we employ the density-dependent expression
$ v(\phi) = v_0 - \zeta\phi$ 
that is widely used in effective hydrodynamic models for MIPS \cite{SBML2014prl,SMBL2015jcp,BiWi2020prr}. Thereby, $v_0$ represents an effective speed of an individual self-propelled particle while $\zeta$ results from the force imbalance related to the self-trapping due to interactions with other particles as frequently occurring at larger densities \footnote{\label{fn:-v0}Note that $v_0$ might become negative as it acquires a negative contribution due to the shift from a proper particle density as in \cite{BiWi2020prr} to the scaled and shifted $\phi$. Then, $\zeta$ remains a positive constant as in a $\rho$-based model. However, here we provide a parameter study in terms of $v_0$ and $\zeta$. Taking into account that symmetries of Eqs.~\eqref{eq:dtphi} and \eqref{eq:dtP} imply identical behavior for $v_0<0, \zeta>0$ and $v_0>0, \zeta<0$, here we focus on the latter case. In other words, $\zeta<0$ ensures a reduced $v(\rho)$ for on average negative $\phi$.}. Note that more complicated expressions (e.g., $v(\phi, \nabla^2\phi)$) may also be employed \cite{CaTa2013el,RaBZ2019epje,BiWi2020prr} but do not amend our central result.

\begin{figure}
\centering
\includegraphics[width=\hsize]{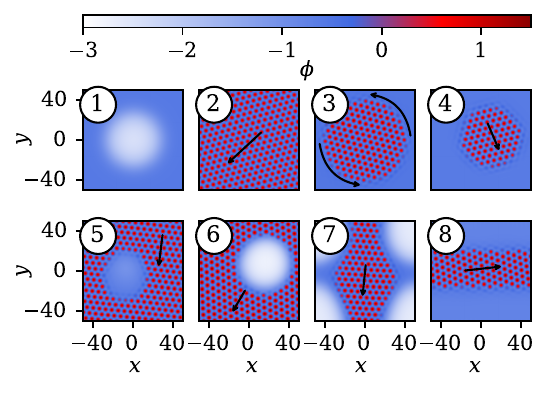}
\caption{Typical snapshots of various resting and moving states are represented by their density profiles $\phi(\vec{x},t)$. Panels  \circlenumber{1} to  \circlenumber{8} show states at parameter values indicated by corresponding numbers in \cref{fig:morphological_PDs}~(a) to (c).
Given are  \circlenumber{1} a phase-separated state of a gas bubble inside a liquid,  \circlenumber{2} an extended traveling crystal,  \circlenumber{3} to  \circlenumber{8} various types of moving crystallites. The black arrows indicate the direction of motion of the crystalline regions.}
\label{fig:solutions}
\end{figure}

The higher-order active PFC model \eqref{eq:dtphi} and \eqref{eq:dtP} shows the full range of behaviors sketched in Fig.~\ref{fig:cluster}: In the passive case ($v=0$) the full phase behavior of a system with gas, liquid and crystalline state is reproduced \cite{WLHD2020prm}, shown here in \cref{fig:morphological_PDs}~(a) by a phase diagram in the plane spanned by mean density $\bar{\phi}$ and effective temperature $T$ with a focus on the vicinity of the triple point at $(\bar\phi, T)\approx (-0.5,0)$. For details on the employed time simulations see section~B of the SM. There is a one-phase region of uniform states (blue, liquid or gas), and two-phase regions of liquid-gas (orange), solid-liquid (dark green), and solid-gas coexistence (light green). The white lines indicate linear stability thresholds for the uniform state, i.e., spinodals, with respect to liquid-gas phase separation (dashed) and crystallization (dotted). They correspond to Cahn-Hilliard (stationary, large-scale) and conserved-Turing instabilities (stationary, small-scale), respectively (see classification of \cite{FrTh2023prl}). Typical dispersion relations and complete spinodals are given in section~C of the SM.

Increasing the density-independent activity $|v_0|$ (at $\zeta=0$) from zero significantly changes the phase diagram (see \cref{fig:morphological_PDs}~(b)). Overall, all clustering is reduced, the region of liquid-gas phase separation has disappeared and crystallization is shifted toward larger densities. The dot-dashed spinodal now indicates the onset of a conserved-wave instability~\cite{FrTh2023prl}, i.e., related to the direct emergence of traveling crystalline states. The shift is consistent with the suppression of crystallization with increasing activity observed in the standard active PFC model \cite{MeLo2013prl,OpGT2018pre,OKGT2020c}. Here, it applies as well to phase separation where with increasing $|v_0|$ the liquid-gas critical point and the spinodals are shifted toward lower temperatures. One further notices in \cref{fig:morphological_PDs}~(b) that crystal-gas coexistence has all but disappeared. The remaining crystalline states all correspond to travelling states, see example states in Fig.~\ref{fig:solutions}. Thereby, finite-size crystallites corresponding to crystal-liquid coexistence frequently rotate as further analyzed below. In summary, pre-existing structures that result from passive forces, i.e., by thermodynamic interactions between particles, are all counteracted by the density-independent activity represented by $v_0$.

An additional density-dependent activity $\zeta$ can reverse the destruction of clusters observed from \cref{fig:morphological_PDs}~(a) to \cref{fig:morphological_PDs}~(b), see \cref{fig:morphological_PDs}~(c) for $v_0=1$ and $\zeta=-0.5$ as well as Fig.~6 of the SM, see note~\cite{Note1}. In particular, the liquid-gas spinodal is restored, and accordingly the liquid-gas critical point is again at higher temperature. In other words, in the corresponding density range the model reproduces the results for MIPS obtained with simpler models that do not account for crystallization. In the resulting region of liquid-gas coexistence resting liquid drops (or gas bubbles) are found, similar to state  \circlenumber{1} in Fig.~\ref{fig:solutions} but now created by motility. Then, in the hatched purple region domain-filling stationarily traveling crystals are found (state  \circlenumber{2}). Coexistence of drops or bubbles with an extended traveling crystal have re-emerged [states  \circlenumber{5} and  \circlenumber{6}, note the liquid wetting layer in the latter]. Finally, state  \circlenumber{7} provides an example of three-phase gas-liquid-solid coexistence in an active system. Note that coexisting states may also exist outside the spinodally unstable region, i.e., the borders between the colored regions in \cref{fig:morphological_PDs} represent finite-size approximations of (non)equilibrium binodals. This indicates that all transitions are of first order as also confirmed by the existence of various localized states~\cite{TARG2013pre, TFEK2019njp, HAGK2021ijam}. A detailed understanding of the intricate relations between and multistability of the various uniform, periodic and localized states (that may rest or travel), in particular, in the vicinity of the triple point can be achieved by considering the bifurcation structure for finite systems. A selection of corresponding results is given in section~D of the SM. Note that in contrast to most results described for the standard active PFC model \cite{MeLo2013prl,OpGT2018pre,OKGT2020c}, here, the group velocity of the crystallites is lower than their phase velocity, i.e., individual peaks move faster than the overall structure [states \circlenumber{4}, \circlenumber{5}, \circlenumber{6}, \circlenumber{7}, \circlenumber{8}) in Fig.~\ref{fig:solutions}]. Physically, this corresponds to particles melting at the front of the traveling crystallite into the coexisting uniform phase and crystalising at the rear. In other words, at the rear particles get stuck in the ``slowly moving traffic jam'' of the crystallite, then slowly move through it until they detach themselves when reaching the front.

\begin{figure}[h]
\centering
\includegraphics[width = 0.9\hsize]{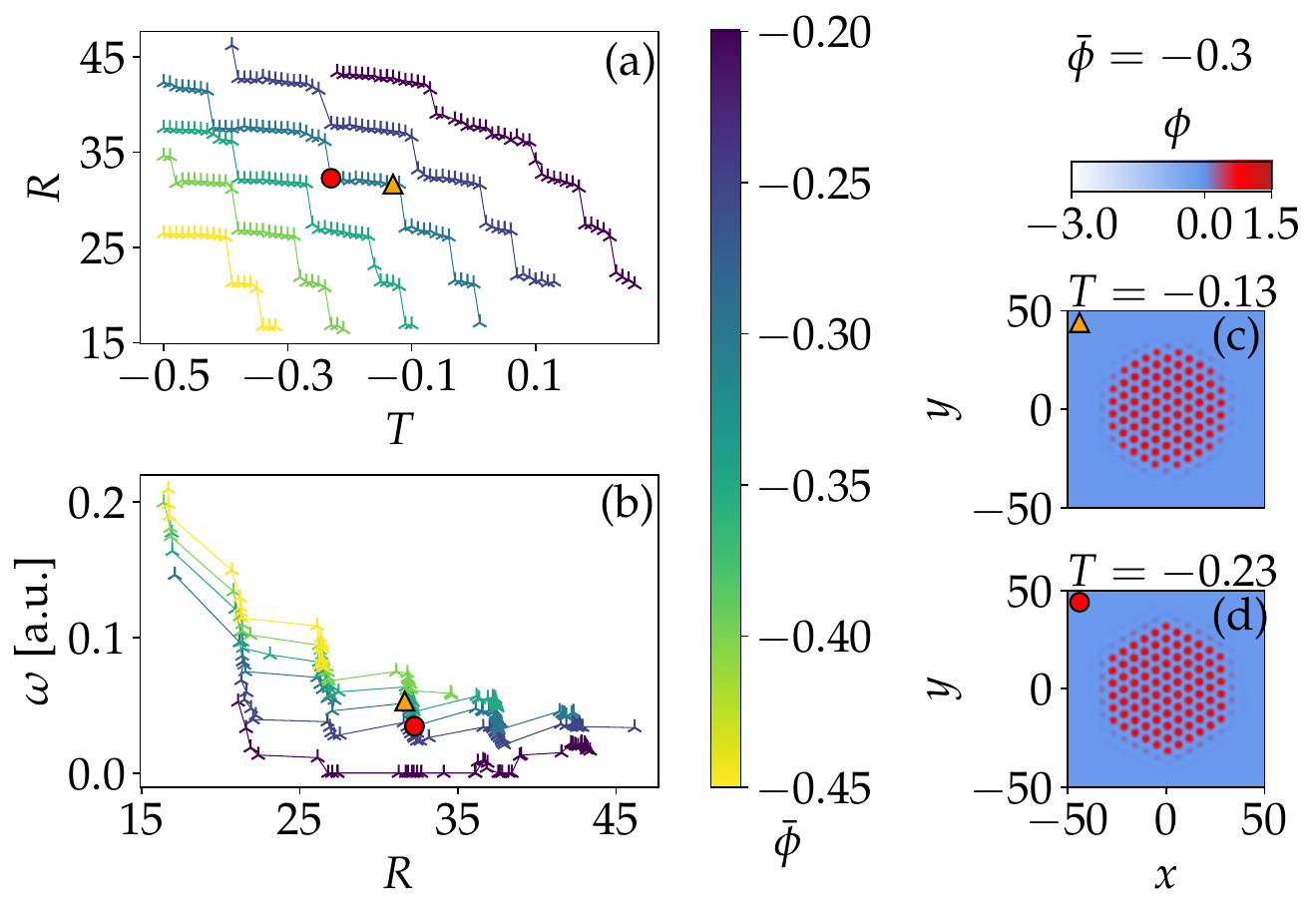}
\caption{Characteristics of rotating crystallites for the case of $v_0=1$ $\zeta=1$. Shown are (a) the radius $R$ in dependence of the temperature $T$ and (b) the angular velocity $\omega$ (in degree per time unit) as a function of $R$, both at several fixed mean densities $\bar{\phi}$ as indicated by the color bar. Panels (c) and (d) show profiles of states at the two ends of a single plateau at $\bar\phi=-0.3$ in (a) where they are marked by an orange triangle and red circle respectively. The domain size is $L_x=L_y=100$, for numerical details see SM Section~B.}
\label{fig:aPFC-r-v-overview}
\end{figure}

\begin{figure}[tb]
\centering
\includegraphics[width = 0.9\linewidth]{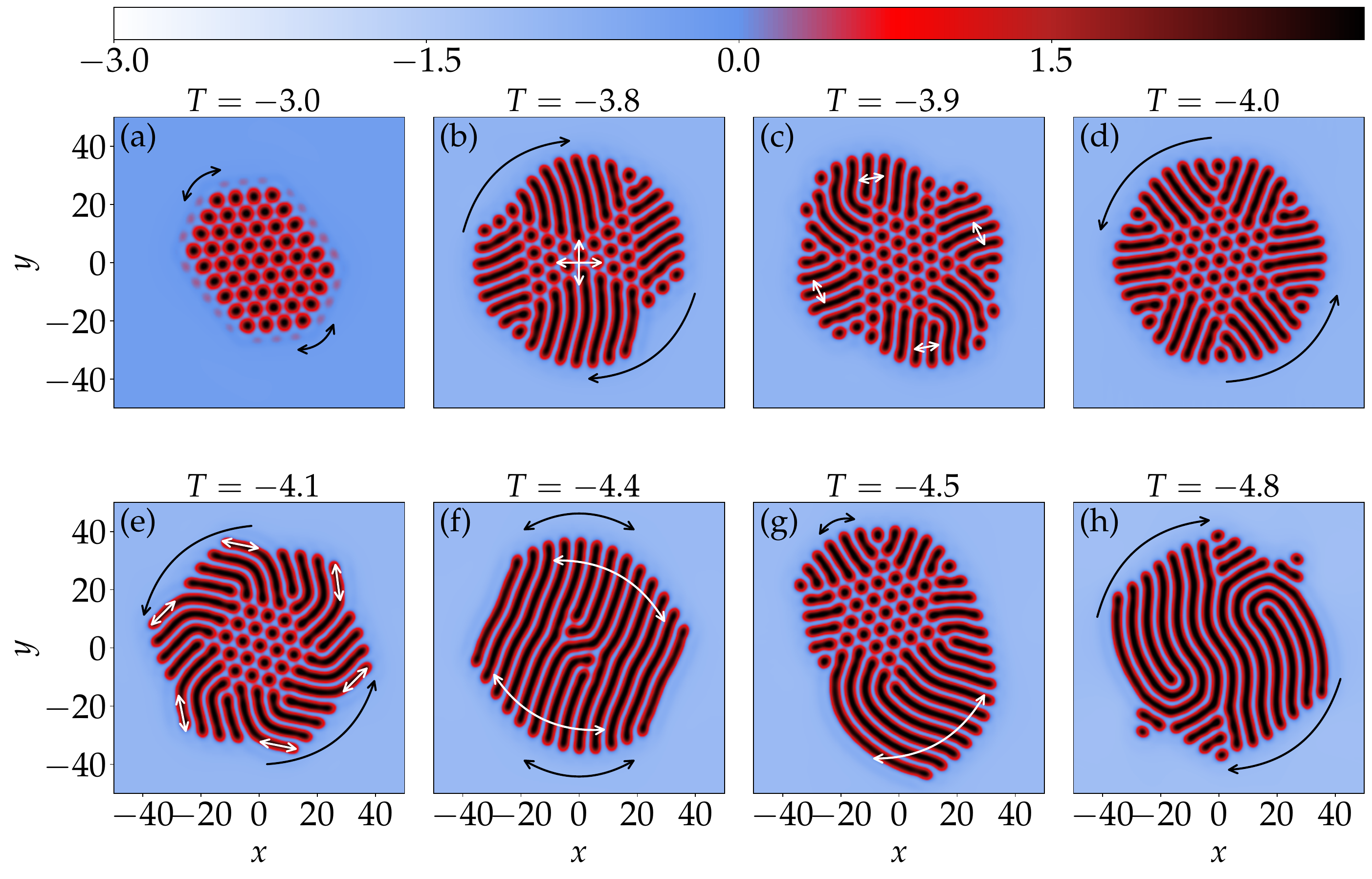}
\caption{Examples of intricate behavior of crystalline clusters at lower $T$ and $\bar\phi=-1$. The black and the white arrows indicate main overall and inner modi of motion, respectively, and $\phi$ is given by the colorbar. The individual states are described in the main text, also cf.~movies in the SM. Remaining parameters are as in Fig.~\ref{fig:aPFC-r-v-overview}.}
\label{fig:far-from-3p}
\end{figure}

An intriguing state are rotating crystallites as state~\circlenumber{3} in Fig.~\ref{fig:solutions}. Such states are described for many experimental systems~\cite{HBCD2024prl,PeWL2015prl,TMLC2022nat,YaBG2015sm} and microscopic simulations~\cite{MaWi2023ps} but are rarely mentioned in studies employing PFC models. Here they frequently occur beside the also found traveling crystallites [state~\circlenumber{4} in Fig.~\ref{fig:solutions}]. The rotating crystalline clusters move as a nearly hexagonal rigid body, i.e., with identical group and phase velocity. Fig.~\ref{fig:aPFC-r-v-overview} provides an overview of their characteristics. Their radius $R$ and angular velocity $\omega$ are given as a function of $T$ and $R$, respectively, at several fixed densities $\Bar{\phi}=-\,0.2\dots -0.45$. The dependence $R(T)$ has the appearance of a staircase that goes up with decreasing $T$ while the steps are getting longer (at equal height), a typical behavior for localized states that show homoclinic snaking~\cite{BuKn2006pre, Knob2016ijam}. Inspecting profiles like Figs.~\ref{fig:aPFC-r-v-overview}~(c) and (d) shows that along each nearly horizontal part an initially incomplete outer shell of the hexagonal structure laterally grows more peaks with decreasing $T$. When a shell is completed, a new outer layer is initiated by peaks in the central part of each face that then grows, thereby resulting in the step-like increase of $R$. In other words, each step corresponds to a stable radius. A corresponding structure is visible in the dependence of $\omega$ on $R$ [Fig.~\ref{fig:aPFC-r-v-overview}~(b)] where for small clusters the velocity decreases with increasing cluster size while at large $R$ one can discern a small increase. Note the for the largest $\bar\phi$ the clusters seem to remain at rest in an intermediate parameter range. Beside the shown clusters of six-fold symmetry, also less symmetric states exist that feature facets of different lengths (not shown). Further note that for the employed domain size, clusters above $R_\text{max}\approx 46$ interact via the boundaries, ultimately form other states, and are therefore excluded from the analysis.
Similarly, there is a minimal radius $R_\text{min}\approx 16$ corresponding to the smallest stable cluster (seven peaks). Lower densities result in uniform states ~\footnote{At the higher mean densities $\bar{\phi}=-0.2$ and $\bar{\phi}=-0.25$, one finds a larger $R_\text{min}\approx 21$ due to the softening of the interface between uniform background and pattern. The interface is quite sharp at low $T$ but widens at larger $T$.}.

At more extreme conditions, e.g., lower temperature and larger density more intricate behavior may occur: There, a stripe phase also exists (in analogy to the standard active PFC model \cite{MeLo2013prl}) that corresponds to particles that are periodically placed in one direction but are mobile in the other one.

This results in the new types of localized states shown in \cref{fig:far-from-3p}. They are often of lower symmetry and either combine local stripe and crystalline arrangements (two-phase cluster) or are pure localized stripe states. Due to activity the entire clusters or their parts may move in different ways. A selection includes an elongated hexagonal crystallite swinging like a torsion pendulum (implying a periodic reversal in the polarization field) in Fig.~\ref{fig:far-from-3p}~(a), an inversion-symmetric two-phase crystallite with a bow-tie shaped crystalline core and outer stripe patches where the core performs a kind of bouncing motion overlaid by a slow rotation and giving the overall impression of wobbling in Fig.~\ref{fig:far-from-3p}~(b), 
a shaking state combining a chiral star-like crystalline core that is nearly at rest and outer striped patches that seem to wave in Fig.~\ref{fig:far-from-3p}~(c),
a similar but slightly more regular structure that rotates in Fig.~\ref{fig:far-from-3p}~(d)
a rotating structure with hexagonal crystalline core covered by chiral stripes that additionally wobble in Fig.~\ref{fig:far-from-3p}~(e), an inversion-symmetric swinging striped crystallite with additional waving stripe motion in Fig.~\ref{fig:far-from-3p}~(f),
a swinging strongly asymmetric two-phase crystallite where the long stripes additionally wave in Fig.~\ref{fig:far-from-3p}~(g),
and a rotating inversion-symmetric striped crystallite in Fig.~\ref{fig:far-from-3p}~(h). 

To conclude, we have presented an active Phase-Field-Crystal (PFC) model capable of describing (active) gas, liquid and crystal phases, including their triple point. With the model we have studied the interplay of passive (i.e., thermodynamic) and active (i.e., motility-induced) condensation, crystallization and phase coexistences employing density-independent and density-dependent activities. This includes a variety of traveling and rotating crystallites that may coexist with gas or liquid as further evidenced by phase diagrams and a detailed study of rotating crystallites.

\nocite{Boyd2001dover}
\nocite{EGUW2019springer}
\nocite{UeWR2014nmma}

\bibliographystyle{unsrturl}


\end{document}


\title{Supplementary Material - Motility-induced crystallization and rotating crystallites}
\author{Max Philipp Holl}
\email{max.holl@aalto.fi}
\thanks{ORCID ID: 0000-0001-6451-9723}
\thanks{M.H. and A.S. contributed equally to this work.}
\affiliation{Department of Chemistry and Materials Science, Aalto University, P.O. Box 16100, FI-00076 Aalto, Finland}
\affiliation{Academy of Finland Center of Excellence in Life-Inspired Hybrid Materials (LIBER), Aalto University, P.O. Box 16100, FI-00076 Aalto, Finland}
\affiliation{Institute of Theoretical Physics, University of M\"unster, Wilhelm-Klemm-Str.\ 9, 48149 M\"unster, Germany}

\author{Alina Barbara Steinberg}
\email{a\_stei52@uni-muenster.de}
\thanks{M.P.H. and A.S. contributed equally to this work.}
\thanks{ORCID ID: 0000-0001-6598-9700; MH and AS contributed equally to this work.}
\affiliation{Institute of Theoretical Physics, University of M\"unster, Wilhelm-Klemm-Str.\ 9, 48149 M\"unster, Germany}

\author{Uwe Thiele}
\email{u.thiele@uni-muenster.de}
\homepage{http://www.uwethiele.de}
\thanks{ORCID ID: 0000-0001-7989-9271}
\affiliation{Institute of Theoretical Physics, University of M\"unster, Wilhelm-Klemm-Str.\ 9, 48149 M\"unster, Germany}
\affiliation{Center for Nonlinear Science (CeNoS), University of M\"unster, Corrensstr.\ 2, 48149 M\"unster, Germany}
\affiliation{Center for Multiscale Theory and Computation (CMTC), University of M\"unster, Corrensstr.\ 40, 48149 M\"unster, Germany}

\maketitle
\section*{Supplementary Material}

The Supplementary Material provides further details and background information regarding the higher-order active Phase-Field-Crystal (PFC) model presented and analyzed in the main text. In particular, section~\ref{app:model} gives details regarding the employed energy functional, and presents the resulting specific kinetic equations with the complete set of parameters. Subsequently, section~\ref{sec:numerics} details the employed numerical methods and related parameters while section~\ref{app:linstab} presents the linear stability analysis of uniform states, discusses the types of resulting dispersion relations and the corresponding spinodals and their dependence on parameters. Finally, section~\ref{sec:1dresults} discusses the intricate bifurcation structure of uniform, periodic, and localized states in the case of one-dimensional domains that in conjunction with the phase diagrams in the main text allows for a deeper understanding of the system behavior.

\subsection{Energy functional, governing equations and parameters}
\label{app:model}
First, we provide details on the employed energy functional and the corresponding specific used parameters. The free energy functional $\mathcal{F}[\phi,\vec{P}]$ underlying the gradient dynamics in the main part combines the density-dependent higher-order free energy $\mathcal{F}_\text{PFC}[\phi]$ developed by Wang \textit{et al.} \cite{WLHD2020prm} and an orientational part $\mathcal{F}_\vec{P}[\vec{P}]$. The latter is chosen identical to the corresponding part of the energy for the standard active PFC model \cite{MeLo2013prl,OKGT2020c,OKGT2021pre}. Overall we have
\begin{align}
\mathcal{F}[\phi,\vec{P}] = \mathcal{F}_\text{PFC}[\phi] + \mathcal{F}_\vec{P}[\vec{P}].
\label{eq:free_energy_combined}
\end{align}
The density-dependent part is
\begin{align}
\mathcal{F}_\text{PFC}[\phi] = \int &-B_0\phi - \frac{1}{2}\phi\left(C_0 + C_2\Delta + C_4\Delta^2 + C_6\Delta^3\right)\phi \notag\\
&- \frac{1}{6}\left[D_0\phi^3+ D_{11}\phi^2\Delta\phi\right] - \frac{1}{24}\left\{E_0\phi^4 + E_{1122}\phi^2[\Delta\phi]^2\right\} \mathrm{d}^nr\\
= \int & f_\text{PFC}\,\mathrm{d}^nr \,.
\end{align}
where $\Delta$ is the Laplace operator and $n$ the spatial dimension. Note that great care is taken in \cite{WLHD2020prm} to develop the specific functional dependencies of the ``parameters'' on an effective (scaled and shifted) temperature $T$ to ensure the functional models a system with gas, liquid and crystal phases and gives a phase diagram of standard form in the plane spanned by $T$ and mean density $\bar\phi$. In particular, $B_0$, $C_0$, $C_2$, and $C_4$ depend on $T$. All parameters and dependencies are given in \cref{tab:parameters}. They are also employed in the present work. The higher-order terms $D_{11}\phi^2(\vec{r})\Delta\phi(\vec{r})$ and $E_{1122}\phi^2(\vec{r})[\Delta\phi(\vec{r})]^2$ stem from three- and four-point correlations, respectively, and are essential to obtain the correct sequence of phase transitions. For nonzero $D_{11}$, one needs $E_{1122} <0$ to prevent divergent large-$k$ behavior at very large mean densities $\bar\phi$.
\begingroup
\setlength{\tabcolsep}{10pt} 
\renewcommand{\arraystretch}{1.5} 
\begin{table*}[ht!]
\centering
\caption{\label{tab:parameters} Parameters and functional depenencies used in the free energy $\mathcal{F}_\text{PFC}[\phi]$ of the passive PFC model as determined in Ref.~\cite{WLHD2020prm}. In particular, $B_0$, $C_0$, $C_2$, and $C_4$ depend on $T$ .}
\begin{tabular}{|c|c|c|c|c|c|c|c|c|}
\hline
$B_0$&$C_0$&$C_2$&$C_4$&$C_6$&$D_0$&$D_{11}$&$E_0$&$E_{1122}$\\ 
\hline
$-4.5 - 3T$ & $-5.764 - T$ & $17.8 + 2T$ & $39.8 - T$ & $16$ & $-9$ & $-34.2$ & $-6$ & $-52.1$\\ 
\hline 
\end{tabular}
\end{table*}
\endgroup

The polarization-dependent part is \cite{MeLo2013prl}
\begin{align}
\mathcal{F}_\vec{P}[\vec{P}] = \int \frac{c_1}{2} \vec{P}^2 + \frac{c_2}{4}\vec{P}^4 \mathrm{d}^nr.
\end{align}
For $c_1<0$ and $c_2>0$ it allows for spontaneous polarization. However, we follow most earlier analyses \cite{MeLo2013prl,OKGT2020c,OKGT2021pre} and simply use $c_1=1$ and $c_2 = 0$.
Introducing the free energy \cref{eq:free_energy_combined} into the kinetic equations Eqs.~(2) and (3) of the main text we obtain 
\begin{align}
\dt\phi = \Delta\biggl\lbrace &-(C_0 + C_2\Delta + C_4\Delta^2 + C_6\Delta^3)\phi-\frac{1}{2}D_0\phi^2 - \frac{1}{6}D_{11}(2\phi\Delta\phi + \Delta(\phi^2)) \notag\\
&\left. -\frac{1}{6}E_0\phi^3 - \frac{1}{12}E_{1122}[\phi(\Delta\phi)^2 + \Delta(\phi^2\Delta\phi)]\right\rbrace-\nabla \cdot[v(\phi) \vec{P}], \label{app:eq:dtphi}\\
\dt \vec{P} = D_c&\Delta\vec{P} - D_{nc}\vec{P} - \nabla\cdot[\alpha v(\phi)\phi].\label{app:eq:dtP}
\end{align}
Further, we set the mobility constants for the conserved and the nonconserved part of the polarization dynamics to $D_c=0.2$ and $D_{nc}=0.5$, respectively. These parts correspond to translational and rotational diffusion of the polarization, respectively. The constant in the coupling term is fixed to $\alpha=0.5$, to be consistent with, e.g., Ref.~\cite{SBML2014prl}.

In the passive limit, \cref{app:eq:dtphi,app:eq:dtP} decouple and steady states (with $\dt\phi = 0$) are obtained by solving the twice integrated \cref{app:eq:dtphi}, i.e.,
\begin{align}
0 = &-(C_0 + C_2\Delta + C_4\Delta^2 + C_6\Delta^3)\phi-\frac{1}{2}D_0\phi^2 - \frac{1}{6}D_{11}(2\phi\Delta\phi + \Delta(\phi^2))\notag \\
& -\frac{1}{6}E_0\phi^3 - \frac{1}{12}E_{1122}[\phi(\Delta\phi)^2 + \Delta(\phi^2\Delta\phi)] - \tilde{\mu}. \label{eq:dtphi_mu}
\end{align}
The integration constant $\tilde{\mu}$ is related to the chemical potential $\mu$ by $\tilde{\mu} = B_0+\mu$. Note that the constant of the first integration is set to zero as there is no flux across the boundaries. Equilibrium steady states described by \eqref{eq:dtphi_mu} may be followed through parameter space employing numerical path continuation while the dynamics described by \eqref{app:eq:dtphi} and \eqref{app:eq:dtP} can be studied employing numerical time integration. Information on the numerical methods is given in the next section.

\subsection{Numerical methods and set-ups}
\label{sec:numerics}
Time simulations are used to create phase diagrams and example profiles of steady and dynamic states (Fig. 1-5 of main text) and determine the features of the rotating crystallites. To that end a pseudo-spectral semi-implicit Euler method is employed \cite{Boyd2001dover}. 

The solution measure used for the phase diagrams in Fig. 2 of the main text is the solution type, which is optained by visual inspection. Solution measures for the rotating crystallites (Fig. 4 of the main text) are the radius and the angular velocity. The former is obtained as a time average by taking a horizontal slice at $y=0$ and then measuring at which point $\phi(x,0)$ first reaches a threshold value $A$, starting from the outer edge and moving in. The threshold value is defined as $A = (\phi_h-\phi_b)p$, i.e., the difference of the highest peak density $\phi_h$ and the background density $\phi_b$ is multiplied by a percentage $p$. For odd localized states we use $p=0.75$. For even localized states we use $p=0.65$ as the lack of a central peak makes for a lower value of $\phi_h$ when averaging over time. The time average of the $x$-value where $A$ is passed gives the cluster radius $R$. The angular velocity is given in radians per time step. All simulations are performed on a quadratic domain with $L_x=L_y=100$ and a discretization of $N_x=N_y=256$ at parameters given in \cref{tab:parameters}.

The simulations in the passive phase diagram (Fig. 2~(a) of the main text) are initialized with a circular patch of radius $R_I = 25$ and density $\phi(\vec{r})-\bar\phi = 2$ on a random background, which is then shifted to match $\bar\phi$. All simulations in the active phase diagrams were initialized with the counterpart states in the passive limit. This corresponds to letting the system evolve towards thermodynamic equilibrium and subsequently bringing it out of equilibrium by introducing activity. Initial condition for all rotating crystallite states in Figs. 4 and 5 of the main text are rotating crystallite states from previous simulations at neighboring parameter values.

To create bifurcation diagrams, numerical path continuation \cite{EGUW2019springer} is used for a one-dimensional domain. To that end the \texttt{Matlab} package \texttt{pde2path} \cite{UeWR2014nmma} is used. On the one hand, we follow individual states in parameter space and obtain the branches of uniform, periodic and localized states in \cref{fig:bd_passive_norm,fig:bd_passive_f,fig:bd_passive_mu_phi_om}. On the other hand, we follow saddle-node bifurcations using two-parameter continuation to obtain the ranges of existence summarized in \cref{fig:fp_phi_dT_v00_zeta0_lambda0}.

Solution measures for the bifurcation diagrams are the mean grand potential $\bar \omega$, the chemical potential $\mu$, the mean concentration $\bar\phi$,  the relative mean free energy density $\bar f = \bar f_\text{PFC}-f_0$, where $f_0$ is $f_\text{PFC}$ for $\phi=\bar\phi$ and the L$_2$-norm
\begin{align}
\lVert\delta\vec{u}\rVert = \sqrt{\frac{1}{L}\int_{-L/2}^{L/2} (\phi-\bar\phi)^2 + P^2\,\mathrm{d}x},\label{eq:L2norm}
\end{align}
where $L$ is the domain size.

Note that bifurcation diagrams that show $\mu$ over $\bar\phi$ are appropriate for a conserved dynamics (i.e., $\bar\phi$ is controlled) and indicate corresponding stability w.r.t.\ mass-conserving perturbations. If, in contrast the same branches of states are shown in a bifurcation diagram that is plotted over $\mu$ the diagram is appropriate for a nonconserved dynamics (i.e., $\mu$ is controlled) and indicates corresponding stability w.r.t.\ perturbations at fixed $\mu$ (that normally change mass). See conclusion of \cite{TARG2013pre}, sections~3.2 and 3.3 of \cite{EGUW2019springer}, and the final part of section~2 of \cite{HAGK2021ijam} for corresponding discussions. Also consider Figs.~4, 6, 9, and 11 of \cite{TFEK2019njp} for corresponding 'turnable' plots in the cases of passive phase separation in 1d and 2d, and crystallization in 1d and 2d, respectively.

\subsection{Linear stability of uniform state - dispersion relations and spinodals}
\label{app:linstab}
Next we provide the linear stability analysis of uniform steady states for the higher-order active PFC model \cref{app:eq:dtphi,app:eq:dtP}. Introducing the notation $\vec{w} = (\phi, P)^\mathrm{T}$, the uniform steady states $\vec{w^*}= (\bar\phi, 0)^\mathrm{T}$ solve Eqs.~\eqref{app:eq:dtphi} and \eqref{app:eq:dtP} with $\dt\vec{w} = \vec{0}$ for any $\bar\phi$ (due to mass conservation). Linearizing the kinetic equations in small perturbations $\delta\vec{w}e^{ikx + \lambda t}$ about $\vec{w^*}$ yields the linear eigenvalue problem
\begin{align}
\lambda\delta\vec{w} = \underline{L}\delta\vec{w}
\end{align}
with the Jacobian 
\begin{align}
\underline{L} = \begin{pmatrix}
L_\phi & -ik(v_0 - \zeta\bar\phi)\\
-ik\alpha(v_0-2\zeta\bar\phi) & L_P
\end{pmatrix}.\label{eq:eaPFC_Jacobian}
\end{align}
Here, 
\begin{align}
L_\phi = &-k^2\biggl[-\left(C_0 + D_0\bar\phi + \frac{E_0}{2}\bar\phi^2\right) + \left(C_2 + \frac{2}{3}D_{11}\bar\phi\right)k^2\notag\\ &- \left(C_4 + \frac{1}{12}E_{1122}\bar\phi^2\right)k^4 + C_6k^6\biggr],\\
L_P = & - (D_ck^2 + D_{nc}).
\end{align}
represent the stability problem in the completely decoupled passive case. In the active case, the two branches of the dispersion relation $\lambda(k)$ are given by
\begin{align}
\lambda_\pm = \frac{1}{2}\left[L_{\phi} + L_P \pm \sqrt{(L_{\phi} + L_P)^2-4\det \underline{L}}\right],
\end{align}
where 
\begin{align}
\det \underline{L}=L_\phi L_P + k^2\alpha[v_0^2 + 2(\zeta\bar\phi)^2 - 3v_0\zeta\bar\phi]
\end{align}
is the determinant of the Jacobian (\ref{eq:eaPFC_Jacobian}). The resulting $\Re(\lambda)$ and $\Im(\lambda)$ correspond to growth/decay rates and frequencies of harmonic modes in dependence of wavenumber $k$. The $\lambda$ are real for $(L_{\phi} + L_P)^2\geq 4\det \underline{L}$. Then, $v_0^2$ and $2(\zeta\bar\phi)^2$ always act stabilizing with increasing $|v_0|$ and $|\zeta|$, as the larger eigenvalue ($\lambda_+$) decreases. Only the term $3v_0\zeta\bar\phi$ can act destabilizing for $\bar\phi<0$, if $v_0$ and $\zeta$ have opposite signs.

\begin{figure}[ht!]
\includegraphics[width=0.9\textwidth]{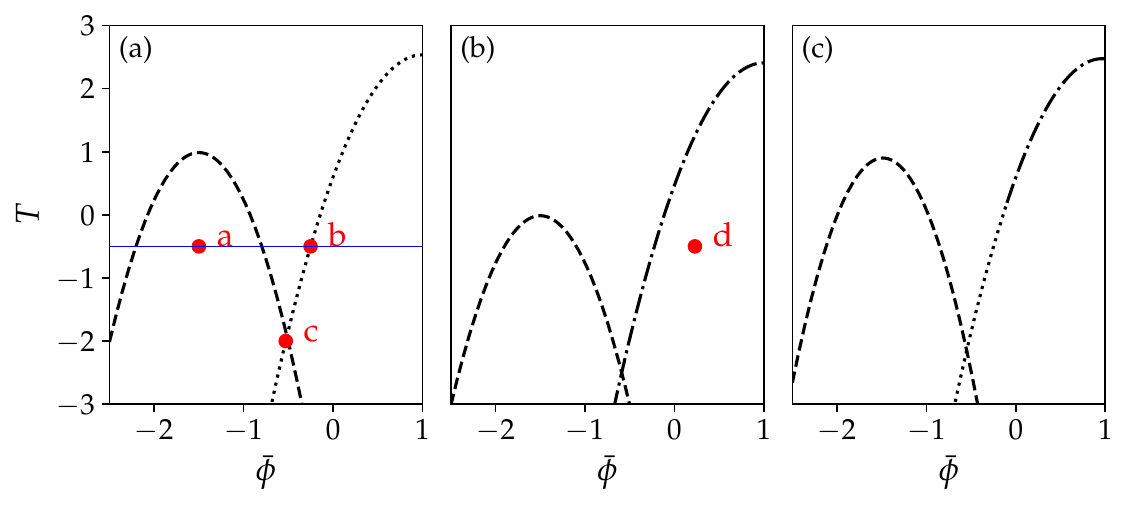}
\caption{Spinodals of the extended PFC model \cref{app:eq:dtphi,app:eq:dtP} in the $(\bar\phi,T)$-plane. Black dashed, dotted, and dash-dotted lines give the spinodals, i.e., the stability tresholds of uniform states with respect to large-scale, stationary (Cahn-Hilliard), small-scale, stationary (conserved-Turing), and small-scale oscillatory (conserved-wave) instability, respectively. The parameters for panels (a)-(c) are identical to the respective parameters for Figs. 2~(a) to (c) of the main text. In particular, (a) represents the passive limit, (b) has $v_0=1$, $\zeta=0$, while (c) shows $v_0=1$, $\zeta=-0.5$. In all three panels there is a codimension-2 point where the onset of large- and small-scale instability coincide, i.e., the spinodals cross. The red dots mark the positions of dispersion relations in \cref{fig:eaPFC_dispersion_relations} and the blue line the parameter choices for \cref{fig:bd_passive_norm}.}
\label{fig:spinodals_map}
\end{figure}

\begin{figure}[ht!]
\includegraphics[width=\textwidth]{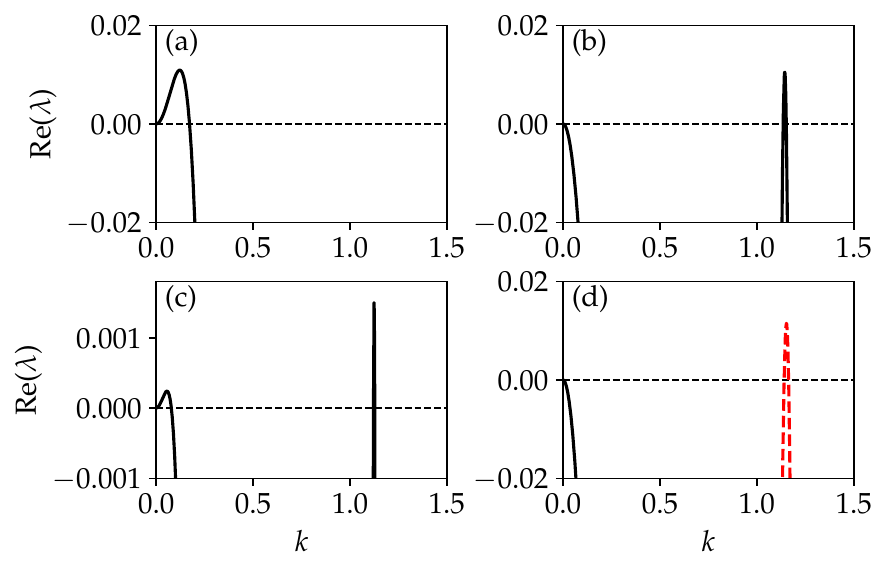}
\caption{Dispersion relations of the extended PFC model in \cref{app:eq:dtphi,app:eq:dtP}. Black solid [red dashed] lines indicate real [complex] eigenvalues.
The dispersion relations presented in (a) at $\bar\phi=-1.5$, $T = -0.5$, and $v_0=0$, (b) at $\bar\phi=-0.2515$, $T = -0.5$, and $v_0=0$, and (c) at $\bar\phi=-0.52895$, $T = -2.$, and $v_0=0$ all show monotonic instabilities. While the dispersion relation in (a) corresponds to a Cahn-Hilliard instability, the one in (b) corresponds to a small-scale (conserved-Turing) instability. Panel (c) presents a case close to the codimension-2 point, i.e., the point where the small-scale and the large-scale stationary instabilities occur simultaneously. Note that in panel (c) the scale of the $y$-axis is much smaller than in the other panels. Panel (d) shows an active case at $\bar\phi =0.229$, $T = -0.5$, and $v_0=1$. The dispersion relation corresponds to an oscillatory small-scale (conserved-wave) instablity.
The remaining parameters are given in \cref{tab:parameters}. }
\label{fig:eaPFC_dispersion_relations}
\end{figure}

The stability thresholds of the uniform state in the $(\bar\phi,T)$-plane are shown in \cref{fig:spinodals_map} for an extended range as compared to Fig. 2 of the main text. The same passive and active cases are shown. In each case, at intermediate $\bar\phi$ liquid-gas phase separation occurs (resulting from a stationary large-scale instability with a conservation law, i.e., a Cahn-Hilliard instability), if the temperature is lower than a critical value $T_c$. At large densities, crystallization occurs due to a stationary small-scale instability with a conservation law, i.e., a conserved-Turing instability in the passive case or due to a small-scale (again conserved-Turing) stationary or small-scale oscillatory (i.e., conserved-wave) instability in the active case. For more information on the employed classification of instabilities, see SM of \cite{FrTh2023prl}. Large- and small-scale instabilities occur simultaneously at the codimension-2 point at $(\bar\phi, T)\approx (-0.5, -2)$. The critical point for phase separation is at $(\bar\phi,T) \approx(-1.5, 1)$ in the passive case. The spinodal line and critical point for phase separation are strongly influenced by density-independent and density-dependent activity. At the chosen sign combination (see corresponding footnote [110] in the main text) the former suppresses phase separation by shifting the critical point and the entire spinodal to lower temperatures while the latter fosters phase separation by reversing the shift. Therefore, the latter represents motility-induced phase separation (MIPS). In contrast, the linear stability threshold for crystallization barely moves for the investigated parameter values. However, the change in the nonlinear behaviour responsible for the coexistence regions outside the linearly unstable regions is more pronounced, see Fig.~2 of the main text and Fig.~\ref{fig:fp_phi_dT_v00_zeta0_lambda0} in the present SM. There density-independent and density-dependent activity suppress and foster crystallization, respectively, indicating that the latter represents motility-induced crystallization (MIC).

Examples for corresponding dispersion relations are presented in \cref{fig:eaPFC_dispersion_relations}. There, black solid and red dashed lines correspond to the real and complex eigenvalues. While in \cref{fig:eaPFC_dispersion_relations}~(a)-(c) dispersion relations in the passive limit are presented, the dispersion relation in \cref{fig:eaPFC_dispersion_relations}~(d) is at $v_0 = 1$. In \cref{fig:eaPFC_dispersion_relations}~(a) at $T = -0.5$, and $\bar\phi = -0.5$ the shown dispersion relation is at parameters well above the onset of a Cahn-Hilliard instability. At the onset of instability, $k=0$ is the mode with the largest growth rate. Above the onset, the eigenvalue at $k=0$ remains zero and there is an adjacent band of unstable wavenumbers. In time simulations, this gives rise to large-scale structures, i.e., a phase-separated state. This instability is similar to the one found for the Cahn-Hilliard equation. At the same temperature $T = -0.5$, but at a higher mean density \mbox{$\phi=-0.2515$} the dispersion relation is at parameters slightly above the onset of a conserved-Turing instability. A small band of wave numbers close to $k\approx1.1444$ is unstable. Directly at onset a time simulation is expected to result in a spatially periodic state, i.e., a crystal. At a codimension-2 point both, the large-scale and the small-scale instability, occur simultaneously. A typcial dispersion relation close to this point is presented in \cref{fig:eaPFC_dispersion_relations}~(c) at $T = -2$ and $\bar\phi = -0.52895$. There, bands of unstable wave numbers exist, one adjacent to $k=0$ and the other around \mbox{$k_c\approx1.1533$}. In a time simulation these two instabilities compete. The nonlinear state that is eventually realized can, however, not be predicted from the linear analysis. Nevertheless, the existence of the codimension-2 point is a first indicator for the existence of three-phase (vapor-liquid-solid) coexistence. Finally, in \cref{fig:eaPFC_dispersion_relations}~(d) we present a dispersion relation for the active case. As in \cref{fig:eaPFC_dispersion_relations}~(b), a small-scale instability occurs, however, here with complex eigenvalues, i.e., the instability is oscillatory and corresponds to a conserved-wave instability. A time simulation produces a traveling periodic state, i.e., a traveling crystal.

\subsection{Bifurcation behavior for one-dimensional states}\label{sec:1dresults}
To gain a deeper understanding of the nonlinear system behavior, we finally present selected bifurcation diagrams of steady states. In particular we focus on the various coexistence regions present in the phase diagram in Fig. 2 of the main text. For relative simplicity and clarity, we limit our attention to one-dimensional systems of a moderate domain size ($L=100$). 

\begin{figure}
\includegraphics[width=0.9\hsize]{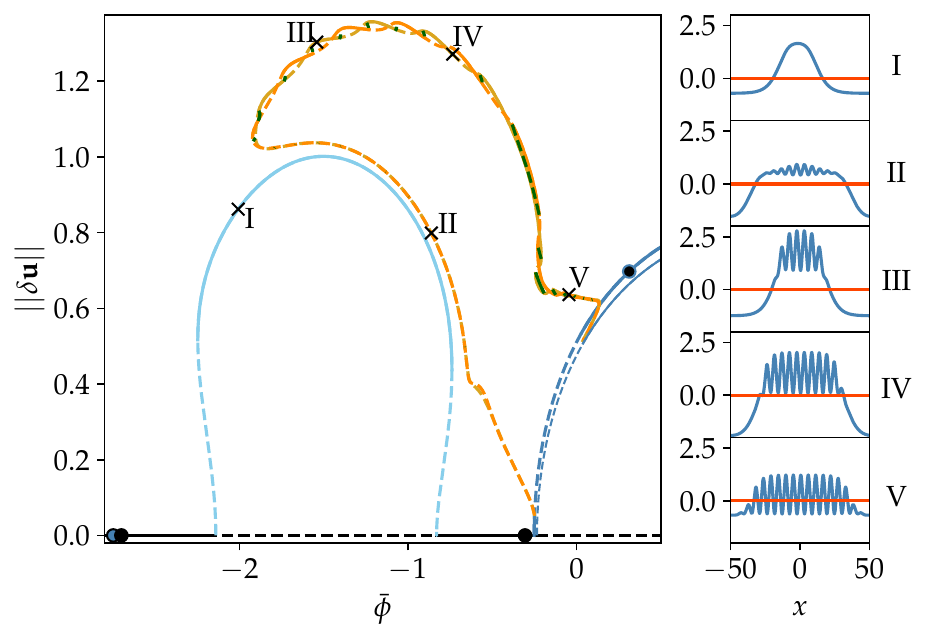}
\caption{\label{fig:bd_passive_norm}Bifurcation diagram for the passive case ($v_0 = 0$, $\zeta=0$) at fixed $T = -0.5$ showing branches of steady states characterized by their L$_2$ norm $\norm{\delta\mathbf{u}}$ as a function of the mean density $\bar\phi$ for a one-dimensional domain of size $L=100$. Solid [dashed] lines indicate linearly stable [unstable] states. The black branch represents homogeneous (gas and liquid) states while the light blue line consists of states of liquid-gas coexistence (phase-separated states). The thick [thin] dark blue line corresponds to domain-filling crystalline (periodic) states with $n=18$ [$n=19$] peaks. The intertwined dark and light orange lines represent the slanted snaking of branches of localized states with odd and even peak number, respectively. They both represent crystal-gas coexistence and are interconnected by branches of asymmetric states (short dark green lines). The four filled circles indicate two pairs of binodal points, where the fill-color corresponds to the color of the branch the state coexists with, e.g., the state marked by the dark blue-filled circle on the black branch of homogeneous states coexists with the crystal state at the black-filled circle on the dark blue branch. Crosses indicate the loci of states I to V whose density profiles $\phi(x)$ are given in the small panels one the right. 
Corresponding free energy densities, grand potential densities and chemical potentials are presented in \cref{fig:bd_passive_f}, \cref{fig:bd_passive_mu_phi_om}~(a) and \cref{fig:bd_passive_mu_phi_om}~(b), respectively. The remaining parameters are as in \cref{tab:parameters}. }
\end{figure} 

\begin{figure}
\includegraphics[width=0.9\hsize]{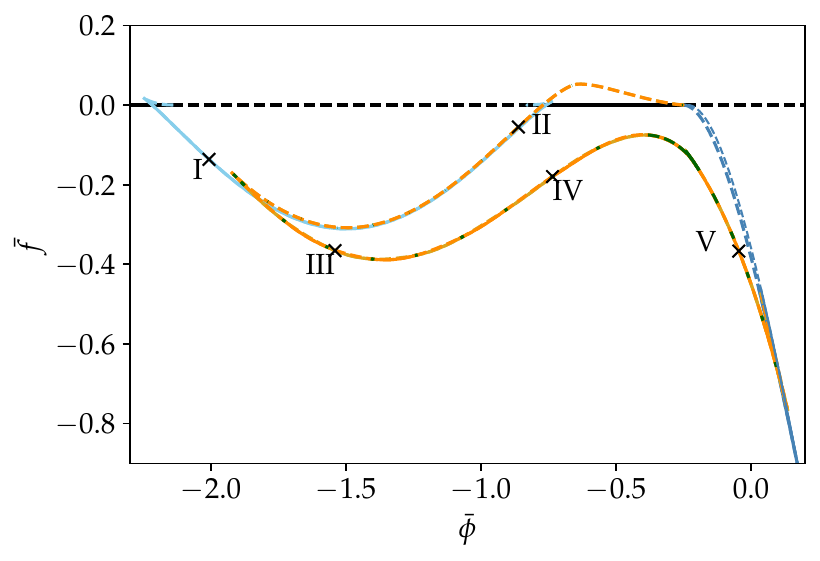}
\caption{\label{fig:bd_passive_f} Shown is the relative mean free energy density $\bar f$ as a function of the mean density $ \bar\phi $ for the passive case analyzed in \cref{fig:bd_passive_norm}. All parameters, line styles and symbols are as in \cref{fig:bd_passive_norm}.}
\end{figure}

\begin{figure}
\includegraphics[width=0.9\hsize]{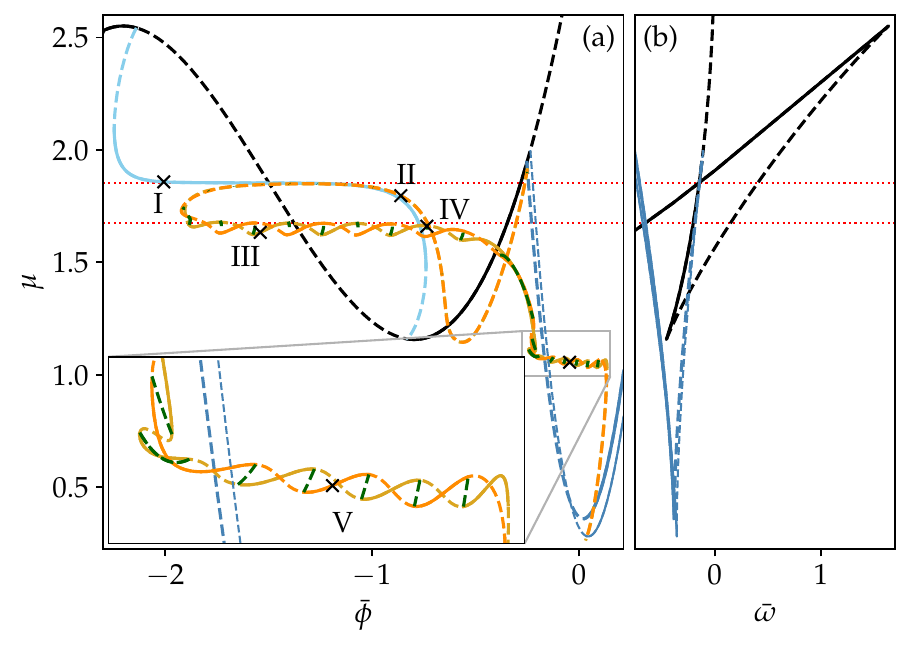}
\caption{\label{fig:bd_passive_mu_phi_om} Panels (a) and (b) give the chemical potential $\mu$ as a function of the mean density $\bar\phi$ and of the mean grand potential $\bar\omega$, respectively, for the passive case in \cref{fig:bd_passive_norm}. In (b), for clarity, only branches of domain-filling states are included. Intersections of branches of stable states (solid lines) indicate Maxwell (binodal) points. The corresponding states coexist in the thermodynamic limit and are marked by filled circles in \cref{fig:bd_passive_norm}. The horizontal red dotted lines indicate the chemical potential of the pairs of binodal points. Note that at $\mu\approx1.2$ there is almost another such intersection in (b) causing the snaking of branches magnified in the inset of (a). The indicated stabilities correspond to control parameter $\bar\phi$, i.e., the stability with respect to perturbations at fixed $\bar\phi$ (and adapting $\mu$). All parameters, line styles and symbols are as in \cref{fig:bd_passive_norm}. }
\end{figure}

First, we consider the dependence on $\bar\phi$ in the passive case at $T = -0.5$, i.e., we analyze a horizontal cut through the phase diagram in Fig. 2~(a) of the main text marked by a horizontal line in \cref{fig:spinodals_map}. To obtain steady states we solve \cref{eq:dtphi_mu} and employ continuation techniques (see section~\ref{sec:numerics} of SM) to follow branches of uniform and nonuniform states. The latter correspond to liquid-gas, solid-gas and solid-liquid coexistence as well as domain-filling crystals. The bifurcation diagram is presented in \cref{fig:bd_passive_norm} in terms of the L$_2$-norm \cref{eq:L2norm}, in \cref{fig:bd_passive_f} in terms of the mean free energy $\bar f$, and in \cref{fig:bd_passive_mu_phi_om}~(a) in terms of the chemical potential $\mu$. To identify coexisting states (see below), \cref{fig:bd_passive_mu_phi_om}~(b) gives all branches in the plane spanned by grand potential density (negative of pressure) $\bar\omega$ and $\mu$.

First, we inspect \cref{fig:bd_passive_norm}: The state of zero norm exists at all $\bar\phi$ and corresponds to a uniform gas (at low $\bar\phi$) or liquid (at larger $\bar\phi$), two states that may coexist (crossing of the black solid line with itself in \cref{fig:bd_passive_mu_phi_om}~(b) at $(\mu, \omega)\approx(1.675,-0.15)$). At low $\bar\phi$ the uniform state is the only, therefore globally stable state. Increasing $\bar\phi$, eventually it becomes unstable in a subcritical pitchfork bifurcation at $\bar\phi \approx -2.140$. It again gains stability at another such bifurcation at $\bar\phi \approx -0.830$ before it destabilizes again at a supercritical pitchfork bifurcation at $\bar\phi \approx -0.251$. The two leftmost bifurcations are connected by a branch of phase-separated states (blue line) that corresponds (where it is stable) to liquid-gas coexistence, a behavior well known from the Cahn-Hilliard model (see, e.g.,\cite{TFEK2019njp}). As the branch emerges subcritically at both ends, the states are initially unstable and gain stability at saddle-node bifurcations at $\bar\phi\approx -2.247$ and $\bar\phi \approx -0.739$, respectively. Along the part between the two saddle-node bifurcations the phase-separated state is linearly stable, for an example profile see panel~I of \cref{fig:bd_passive_norm}. The corresponding dependencies of the energy on $\bar\phi$ in \cref{fig:bd_passive_f} show that the phase-separated state represents the global energy minimum shortly after passing the left saddle-node bifurcation (above $\bar\phi\approx -2.218$) till $\bar\phi\approx -1.829$ shortly before reaching the state of maximal norm. The latter aspect differs from a simple CH model as, here, beyond this point the global minimum is related to gas/liquid-crystal coexistence (see below). The unstable part of the branch consists of nucleation solutions, i.e., threshold states that have to be overcome for transitions between uniform and phase-separated state, i.e., between metastable and stable state, in the binodal region outside the spinodal.

The Maxwell points of coexistence in the thermodynamic limit are obtained using the continuation procedure described in Ref.~\cite{HoAT2021jpcm}. For liquid-gas coexistence they are indicated in \cref{fig:bd_passive_norm} by the filled black circles on the branch of uniform states. In \cref{fig:bd_passive_mu_phi_om}~(a) the Maxwell line connecting the two points lies on the upper thin dotted horizontal line (at $\mu\approx 1.854$). This line represents the thermodynamic limit and is approached by the central part of the blue branch of phase-separated states. This is expected in accordance with analyses of the relation between bifurcation diagrams for finite domains and phase diagrams (thermodynamic limit) performed for Cahn-Hilliard and PFC models in Ref.~\cite{TFEK2019njp}. Increasing the domain size would in \cref{fig:bd_passive_mu_phi_om}~(a) result in a nearly horizontal branch of stable phase-separated states limited by saddle-node bifurcations that approach the binodal values. This confirms also for the present model that the branch of phase-separated states corresponds to a finite-domain representation of the gas-liquid coexistence and may be taken as an indication of a first order transition.

At the supercritical bifurcation at $\bar\phi\approx-0.251$ a branch of domain-filling periodic states (with $n=18$ peaks, dark blue line) emerges from the uniform state. However, shortly thereafter, another supercritical pitchfork bifurcation at $\bar\phi \approx -0.239$ results in another branch of domain-filling periodic states, this time with $n=19$ peaks. The $n=18$ branch is initially stable but is already at $\bar\phi \approx -0.249$ destabilized in a secondary pitchfork bifurcation. There, two branches of localized states (LSs) emerge subcritically -- one with an odd (LS$_\text{odd}$) and one with an even (LS$_\text{even}$) number of peaks (orange lines). Both are initially unstable, at first closely approach and then follow the branch of the stable phase-separated states till beyond its maximum. The corresponding profiles resemble phase-separated states with an additional distinct small-scale periodic modulation of the high-density plateau. In other words, a patch of weakly crystalline state coexists with a gaseous background, for an example see panel~II of \cref{fig:bd_passive_norm}. Both branches of LSs fold back toward larger $\bar\phi$ in saddle-node bifurcations at $\bar\phi\approx -1.9$. There, the LS$_\text{odd}$-branch is stabilized, while the LS$_\text{even}$-branch remains unstable (one unstable eigenvalue) till it is stabilized in a further pitchfork bifurcation, where a short branch of asymmetric LSs emerges subcritically (first short green branch). A typical asymmetric state can be seen in panel~IV of \cref{fig:bd_passive_norm}. In total, 27 such branches connect the LS$_\text{odd}$-branch and the LS$_\text{even}$-branch together forming snake-and-ladder structures of slanted homoclinic snaking typical for systems with a conservation law, i.e., when the mean density is employed as control parameter \cite{TARG2013pre,Knob2016ijam,HAGK2021ijam}. In contrast, if the chemical potential were used as control parameter the snaking would become vertically aligned (cf.~\cref{fig:bd_passive_mu_phi_om}~(a) when rotated by 90 degree and the discussions in \cite{TARG2013pre,HAGK2021ijam}).

Inspection of the energies in \cref{fig:bd_passive_f} shows that the two branches of symmetric LSs (that correspond to gas-solid coexistence) alternatingly form the global minimum between $\bar\phi\approx -1.829$ and $\bar\phi\approx 0.103$. In this range there is always at least one stable LS and the two branches of symmetric LSs exchange stability via the asymmetric runge states. Further, it is discernible that the energy of the subcritical, i.e., unstable part of the branches of symmetric LS (the orange lines close to the blue lines in \cref{fig:bd_passive_norm}) is slightly larger than the energy of the linearly stable phase-separated states. This indicates that the former represent the threshold states that have to be overcome to reach the stable LS of lowest energy from the metastable phase-separated state. While panels III and IV of \cref{fig:bd_passive_norm} are relatively close to solid-liquid-gas coxistence (in 1D the triple point is at $(\bar\phi, T)\approx(-0.379, -0.271)$) and show ``liquid shoulders'' between the gas and the crystal phase, state~V is a clear liquid-solid coexistence.

\cref{fig:bd_passive_mu_phi_om}~(a) actually shows two distinct snaking structures, one at $\mu\approx 1.676$ [related to the just discussed gas-solid coexistence that is indicated by the crossing of the black and blue solid lines in \cref{fig:bd_passive_mu_phi_om}~(b)], and another one at $\mu\approx 1.06$ [magnified in the inset of \cref{fig:bd_passive_mu_phi_om}~(a)], related to the near liquid-solid coexistence that is in \cref{fig:bd_passive_mu_phi_om}~(b) indicated by the near touching of the saddle-node bifurcation where solid and dashed black lines meet, and the blue solid line. One may say that this liquid-solid coexistence is due to a ``ghost-binodal'' because the branches come very close to each other without actually crossing. Coexistence is indicated by thin horizontal lines in \cref{fig:bd_passive_mu_phi_om}~(a) that are decorated by horizontally aligned branches and snaking structures. In \cref{fig:bd_passive_norm} the second snaking structure occurs in the vicinity of the locus of profile~V. Note that the two parts of the branches of LSs where the two snaking structures respectively occur are separated by a further pair of saddle-node bifurcations on the LS$_\text{odd}$-branch [LS$_\text{even}$]-branch at $\bar\phi\approx-0.218$ and $\bar\phi\approx-0.230$ [$\bar\phi\approx-0.210$ and $\bar\phi\approx -0.243$]. Finally, when the LSs have filled the entire finite domain the branches of LSs end in pitchfork bifurcations on the branch of domain-filling crystal states with $n=19$ peaks. At large densities above $\bar\phi\approx 0.103$, the domain-filling crystal with $n=19$ corresponds to the state of lowest energy (\cref{fig:bd_passive_f}). Overall, one may say that the bifurcation structure combines structures known from CH and PFC models in a way as one would expect in the vicinity of a gas-liquid-crystal triple point, and also shows the expected relation to the phase diagram.

\begin{figure}
\includegraphics[width=\linewidth]{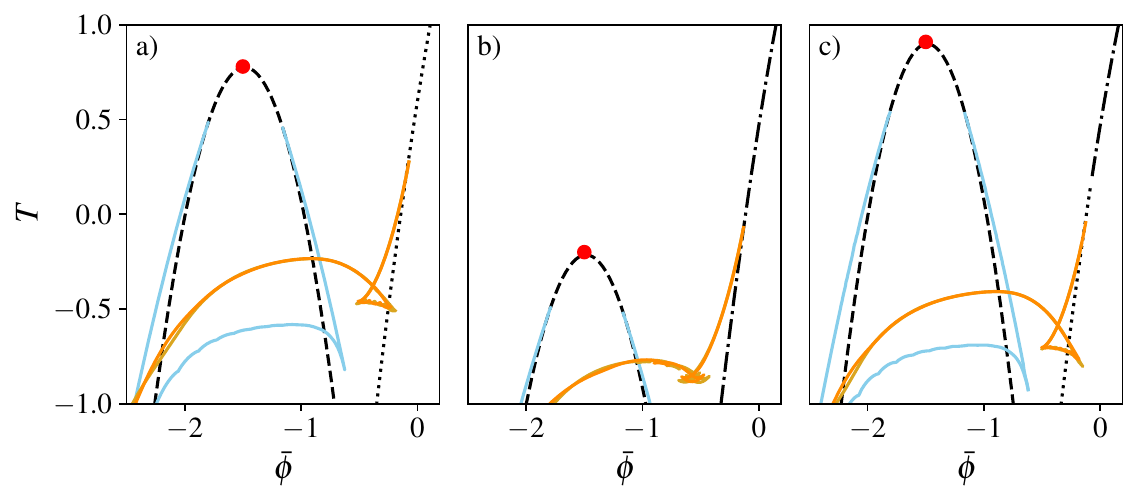}
\caption{\label{fig:fp_phi_dT_v00_zeta0_lambda0}Loci of saddle-node bifurcations in the PFC model in \cref{app:eq:dtphi,app:eq:dtP} at a) $v_0 = 0$, $\zeta=0$, b) $v_0 = 1$, $\zeta=0$, c) $v_0 = 1$, $\zeta=-0.5$. It is always $\lambda=0$. Shown are the loci resulting from two-parameter continuations in $T$ and $\bar\phi$. In case of the yellow and orange branches only the outermost folds for stationary states are tracked. Black dashed and dotted lines denote the spinodal lines for Cahn-Hilliard and conserved-Turing instabilites respectively, while the dot-dashed lines indicate a conserved-wave instability. The red dot marks the crtical point of the gas-liquid spinodal.
The remaining parameters as in \cref{tab:parameters}, while the domain size and colors are is as in \cref{fig:bd_passive_norm}}
\end{figure}

Finally, we briefly discuss how main features of the bifurcation structure change when including activity. To that end we revisit \cref{fig:spinodals_map} for a system of finite size L = 100, i.e., the spinodals are slightly shifted, see \cref{fig:fp_phi_dT_v00_zeta0_lambda0}. Additionally, we track the outermost saddle-node (and primary pitchfork) bifurcations of branches of steady phase-separated and localized crystallite states in the $(\bar\phi,T)$-plane. The loci for the gas-liquid coexistence are shown in light blue and branch off from the spinodal representing the Cahn-Hilliard instability. The remaining colored lines represent the loci of the outermost saddle-node bifurcations for the branches of localized states of odd (orange) and even (yellow) number of peaks. They are nearly on top of each other and represent the complexities of the bifurcation structure as discussed above. Additional traveling states exist in (b) and (c) (not shown) but do not have a significant impact on the overall size of the coexistence region. The swallow tail structures mark the region where the triple point is located. The different parts of the orange line are related to the different coexistences with the crystal state (including metastable pairings). 

Overall, adding the density-independent activity lowers not only the gas-liquid spinodal (critical point marked by the red dot) and associated binodals, but also the binodal related to crystallization. Adding a density-dependent activity, however, counters this effect, and moves up the gas-liquid spinodal and binodals, but also (although to a lesser extent) the liquid-solid binodal. This corresponds to motility-induced phase separation (MIPS) and motility-induced crystallization (MIC).

\bibliographystyle{unsrturl}
